\documentclass{article}      
\usepackage{graphicx}
\usepackage{epstopdf}
\title{Electromagnetic form factors in the Hypercentral Constituent Quark
Model}  

\author{M. De ~Sanctis\\
Istituto Nazionale di Fisica Nucleare, Sezione di Roma,\\
P.le A. Moro 2, 
I-00185 Roma, Italy\\
M.M. Giannini,  E. Santopinto, A. Vassallo\\
Dipartimento di Fisica dell'Universit\`a di Genova\\
and \\
Istituto Nazionale di Fisica Nucleare, Sezione di Genova,\\
via Dodecaneso 33,  
I-16146 Genova, Italy
}

\date{}
\begin{document}

\maketitle  

\begin{abstract}
We discuss the recent experimental results on the ratio bertween the
electric and magnetic proton form factors and how they can be
described by the hypercentral Constituent Quark Model.
\end{abstract}

\section{INTRODUCTION}

 The interest in the electromagnetic form factors of the nucleon has been 
increased by the recent results of the Jefferson Laboratory on the
ratio
between the electric and magnetic form factors of the proton \cite{ped,gay}. At
variance
with the expectations, the ratio
deviates
strongly from $1$ and, for $Q^2~\geq~1~(GeV/c)^2$, the
ratio decreases with an almost linear behaviour, pointing towards the
possible
existence of a zero at $Q^2\approx ~8~(GeV/c)^2$.

These unexpected results pose some problems. The first one is the
compatibility of
the new data with the traditional ones obtained from a Rosenbluth plot.
In this respect much attention is been devoted to the two-photon exchange mechanisms
{twoph},
which however seem to be too small for reconciling the two sets of data. A
critical
re-analysis of the Rosenbluth procedure is also being performed
\cite{ros},
with promising results.

The main further problem is the physical picture emerging from the data, that is the
origin of the decrease of the ratio and the eventual presence of a
zero in the
electric form factor forces.

The planned experiments at higher $Q^2$ will provide the answer about the
occurrence
of a zero in the electric form factor of the proton. From the theoretical
point of
view such zero is a challenge for most theoretical models for the
internal
proton structure. 

In this contribution we report the results of
recent relativistic calculations of the elstic nucleon form factors within the
hypercentral Constituent Quark Model (hCQM).

\section{THE HYPERCENTRAL CONSTITUENT QUARK MODEL}

 In the hCQM the $SU(6)$ invariant quark potential is assumed to be \cite{pl}

\begin{equation}\label{eq:pot}
V(x)÷=÷-frac{1}{x}÷+÷\alpha÷x
\end{equation}
\noindent where $x$ is the hyperradius.
\begin{equation}
x=\sqrt{{\rho}^2+{\lambda}^2},
\end{equation}
\noindent Interactions of the 
type linear plus Coulomb-like have been used since
long time for the meson sector, e.g. the Cornell potential. This form has
been obtained in recent Lattice QCD calculations \cite{bali1,bali2} for
$SU(3)$ invariant static quark sources.
                                       
The  three-quark potential \ref{eq:pot}, depending on the hyperradius
$x$ only,
is hypercentral. Then it can be considered as the lattice
two-body interaction within the so called hypercentral approximation, that is 
averaged over the hyperangle
$\xi=arctg(\frac{{\rho}}{{\lambda}})$ and the
angles $\Omega_{\rho}$, $\Omega_\lambda$; this approximation has been
shown to be
valid, specially for the lower energy states \cite{hca}. On the other
hand, the
hyperradius $x$ depends on the coordinates of all the three quarks,
therefore the interaction $V(x)$ is in general a three-body
potential.

The 'hypercoulomb' part of $-\frac{\tau}{x}$ of the potential \ref{eq:pot} 
 has interesting
properties \cite{hca,sig}. It leads to
a power-law behaviour of the proton form factor and of all the
transition
form factors and it has a perfect degeneracy between
the first $0^+$ excitated state and the first $1^-$ states.

The $SU(6)$ violation is taking into account by adding a
standard
hyperfine interaction $H_{hyp}$ \cite{ik}, treated as a perturbation. The three quark
hamiltonian for the hCQM is then \cite{pl}
\begin{equation}
H÷=÷\frac{p_{\rho}^2}{2m}÷+\frac{p_{\lambda}^2}{2m}÷-÷\frac{\tau}{x}÷
+÷\alpha÷x÷+÷H_{hyp}
\end{equation}

The spectrum is described with $\tau~=~4.59$
and $\alpha~=~1.61~fm^{-2}$ and the standard strength of the hyperfine
interaction needed for the $N-\Delta$ mass difference \cite{ik}.

\section{THE ELECTROMAGNETIC FORM FACTORS}

The model has been used for the prediction of
various physical
quantities of interest, namely the photocouplings \cite{aie},
the electromagnetic transition amplitudes \cite{aie2}, the elastic nucleon
form factors \cite{mds} and the ratio between the electric and magnetic
proton form factors \cite{rap}. 

The calculated r.m.s.
radius of the proton, corresponding to the potential parameters quoted at
the end of the last section, is about $0.5 fm$. This is actually the value
which in earlier calculations has been fitted in order to reproduce the
$D_{13}$ photocouplings \cite{cko,ki}, therefore it is not surprising that the hCQM predicts
helicity amplitudes for the negative parity resonances in fair agreement with data. However,
this low value prevents from a good description of the elastic form factors of the nucleon.

The hCQM is non relativistic and one could in principle think
that this is the origin of the above discussed discrepancies. 
Actually, first order relativistic corrections have been introduced in the
hCQM. The three quark nucleon states have been
boosted to the Breit frame and the matrix elements of the three
quark current have been expanded up to first order in the quark momentum.
The result is that the
theoretical curves are much closer to the experimental data . The most
remarkable effect is, however, obtained for the ratio of the electric ($G_E^p$) and magnetic
($G_M^p$) proton form factors
\begin{equation}
R~=~\mu_{p}~\frac{G_E^p}{G_M^p}
\end{equation}
\noindent where $\mu_p$ is the proton magnetic moment.  The non relativistic
calculations predict the value $R=1$  and introducing the
hyperfine interaction makes no difference ($R=0.99$). However,
the first order relativistic corrections \cite{rap} give rise to a ratio
which is significantly deviating from $1$. It is
interesting to note that the hCQM 
results coincide with the dispersion relation calculation of the Mainz
group \cite{ham}.

Relativity is then a fundamental ingredient for the description of the
elastic nucleon form factors within the hCQM and therefore we have
recently reformulated the model and calculated the elastic
nucleon form factors in a completely relativistic frame. First, we have
introduced in the hCQM the correct relativistic kinetic energy and, using
the same hypercentral potential of Eq. (\ref{eq:pot}), we have obtained an
equivalently good description of the baryon spectrum \cite{rel}. As for
the calculation of the form factors \cite{mds3}, after having  boosted the
new three
quark states to the Breit frame, we have taken into account the quark
current up to any order. Finally, considering that constituent quark may
acquire a finite size, we have introduced quark form factors. The
free parameters in the quark form factors have been used to fit four
set of experimental data, namely the ratio R, the proton magnetic
form factor $G_M^p$, the neutron electric ($G_E^n$) and magnetic
($G_M^n$) form factors \cite{mds3}. The results
for the ratio R are shown in Fig. 1 and are remarkable, since the free
parameters provided by the quark form factors are not sufficient by
themselves to obtain a good fit, it is necessary that already the
pointlike calculations provide a realistic description.
In any case the
size of the quarks obtained in this fit is not larger than $0.3 fm$. It
should be reminded that a recent analysis of the inelastic proton
structure functions \cite{psr} has
shown a possible evidence of the proton containing extended objects with a
size of about $0.2 - 0.3 fm$.

\begin{figure}[ht]
\includegraphics[width=5in]{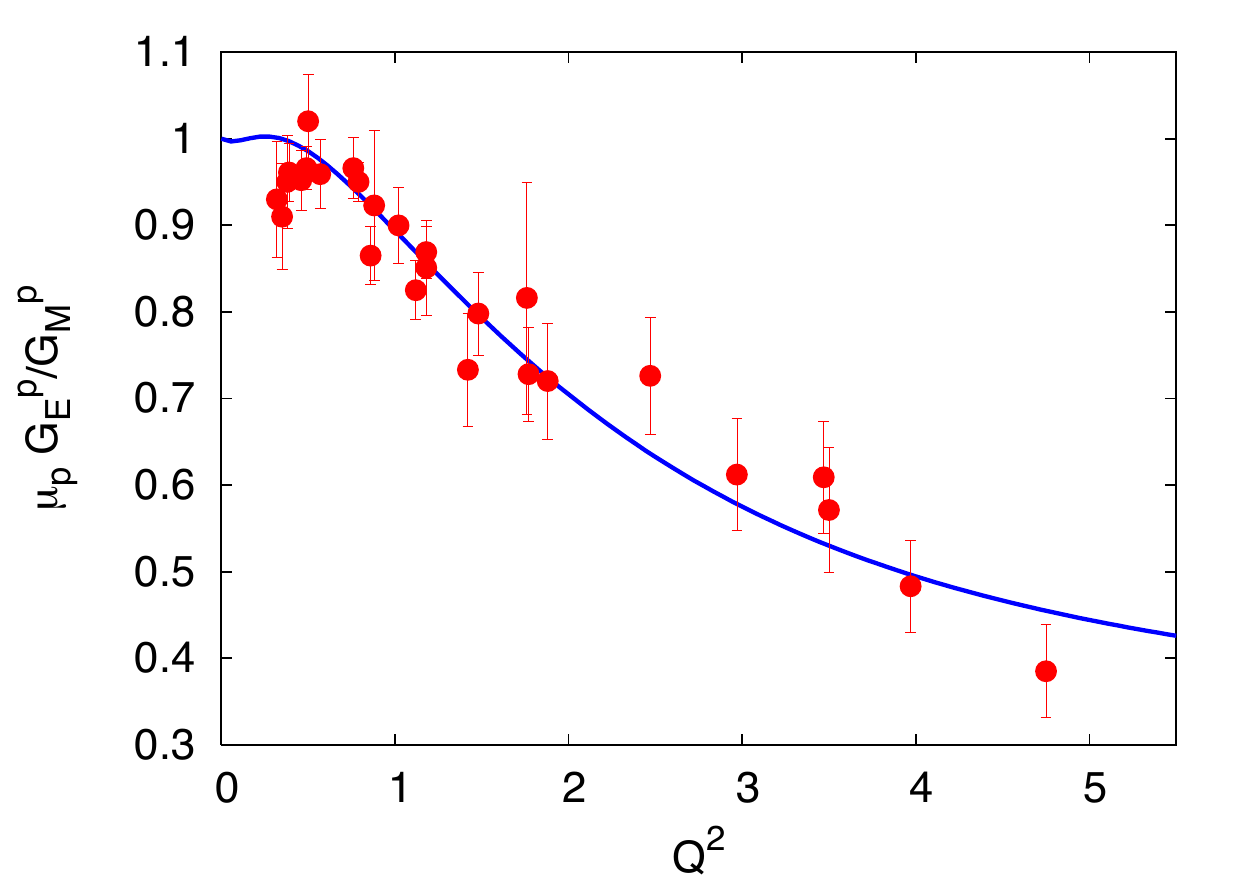}
\caption{ The ratio $\mu_{p}~\frac{G_E^p}{G_M^p}$ calculated with the relativized hypercentral Constituent Quark Model \cite{mds3} (full curve). The data are taken from \cite{ped,gay}.}
\end{figure}

\section{CONCLUSIONS}

The recent Jlab data on the ratio $R~=~\mu_{p}~\frac{G_E^p}{G_M^p}$ show a
strong deviation from the value $1$ predicted by the previous widely
accepted dipole fit and by most models for the internal structure of the
proton. Moreover, extrapolating their trend at higher $Q^2$, one can infer
the presence of a dip. Of course, if the electric form factor of the 
proton has somewhere a zero, then the ratio $R$ is forced to decrease. In
any case the data show that the electric and magnetic distributions of the
proton are quite different. Moreover, as mentioned in the Introduction,
there is the problem of reconciling these new data with those obtained
from a Rosenbluth plot.

The results obtained with the hCQM allow to state that relativity is 
crucial in explaining the decrease of the $R$. It remains to explain the
eventual zero in the proton electric form factor. 

The answer to the question if there is a dip in the proton form factor is 
very important for the understanding of the internal nucleon structure and
it will be hopefully obtained by the planned experiments.

\end{document}